\documentclass{article} 
\usepackage{iclr2026_conference,times}

\usepackage{amsmath,amsfonts,bm}

\def\eqref#1{equation~\ref{#1}}

\def\1{\bm{1}}

\DeclareMathAlphabet{\mathsfit}{\encodingdefault}{\sfdefault}{m}{sl}
\SetMathAlphabet{\mathsfit}{bold}{\encodingdefault}{\sfdefault}{bx}{n}

\usepackage{hyperref}
\usepackage{url}
\usepackage{graphicx}
\usepackage{booktabs}
\usepackage{caption}
\usepackage{xcolor}

\hypersetup{
    colorlinks=true,       
    linkcolor=blue,        
    urlcolor=magenta,         
}

\title{LATO.2: Factorized 3D Mesh Generation with Vertex and Topology Flow}

\iclrfinalcopy
 
\author{
    \normalfont
    \begin{tabular}{@{}l@{}}
    \textbf{Hang Long}$^{1,2,*}$ \hspace{1em}
    \textbf{Tianhao Zhao}$^{1,2,*}$ \hspace{1em}
    \textbf{Junkai Lin}$^{1,2}$ \hspace{1em}
    \textbf{Youjia Zhang}$^{1,2}$ \hspace{1em}
    \textbf{Huipeng Guo}$^{1}$ \\
    \textbf{Rendong Liang}$^{2}$ \hspace{0.1em}
    \textbf{Jiale Xu}$^{2}$ \hspace{0.1em}
    \textbf{Jozef Hladký}$^{3}$ \hspace{0.1em}
    \textbf{Matthias Nießner}$^{4}$ \hspace{0.1em}
    \textbf{Yuanming Hu}$^{2}$ \hspace{0.1em}
    \textbf{Wei Yang}$^{1,\dagger}$ \\
    $^{1}$Huazhong University of Science and Technology \quad
    $^{2}$Meshy AI \quad \\
    $^{3}$Independent Researcher \quad
    $^{4}$Technical University of Munich \\
    \\
    \multicolumn{1}{c}{
        \href{https://github.com/LoHhhha/LATO.2}{https://github.com/LoHhhha/LATO.2}
    }
    \end{tabular}
}

\newcommand{\methodname}{\textsc{LATO.2}}
\begin{document}

\maketitle
\footnotetext[1]{ \boldsymbol{$*$} Equal contribution. \textbf{1,2} This work is done while interning with Meshy AI.}
\footnotetext[2]{ \boldsymbol{$\dagger$} Corresponding author: weiyangcs@hust.edu.cn.}

\vspace{-16pt}

\begin{center}
    \centering
    \captionsetup{type=figure}
    \includegraphics[width=1.0\linewidth]{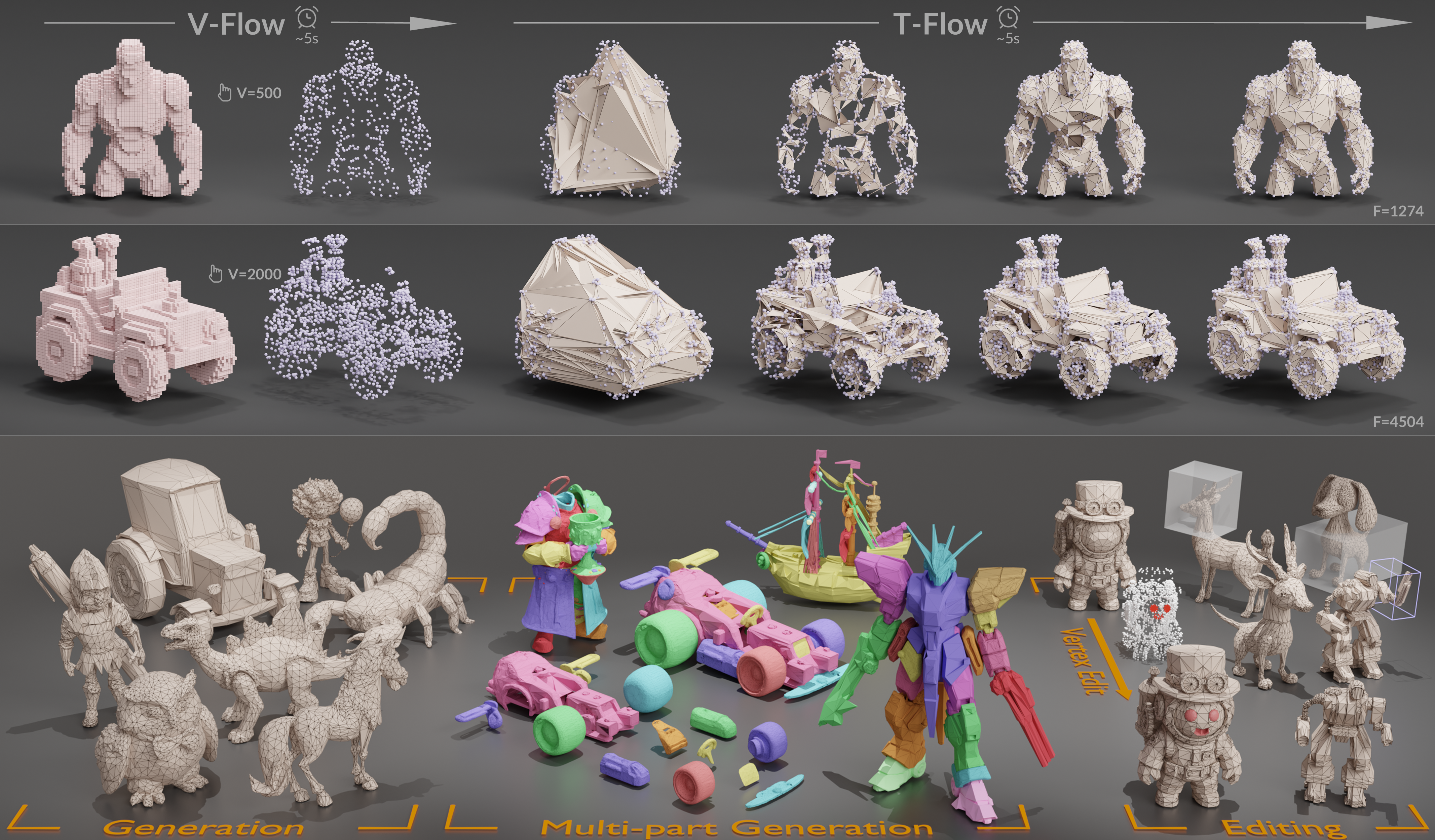}
    \caption{
    We present \methodname{}, which factorizes mesh generation into a vertex flow (V-Flow) generating vertex positions under a controllable vertex count, and a topology flow (T-Flow) predicting connectivity from realized vertices. It supports high-quality generation (bottom left), part-wise generation at scalable resolution (bottom middle), and topology-adaptive editing (bottom right).}
    \label{fig:teaser}
\end{center}%

\begin{abstract}
Flow matching over carefully designed latent representations has recently emerged as a powerful paradigm for topology-aware mesh generation. Existing approaches, however, model vertices and connectivity jointly in a joint latent space, entangling continuous vertex geometry with discrete combinatorial structure; this complicates flow learning and manifests as drifting vertices and broken surfaces. We present \textbf{\methodname{}}, a factorized flow matching framework that decomposes mesh generation into a vertex flow followed by a connectivity flow conditioned on the realized vertices, with both stages anchored to a shared coarse voxel scaffold. Dedicated VAEs underpin the two stages, recovering vertices at sub-voxel precision and embedding discrete connectivity into a continuous latent space. We demonstrate two advantages unique to this factorization:
(i) part-wise generation, in which the scaffold is partitioned and each part synthesized at full latent capacity, yielding substantially
higher-resolution meshes than a monolithic latent permits; and (ii) topology-adaptive editing, in which manipulating first-stage vertices induces
the corresponding connectivity without re-optimization. Experiments show that \methodname{} surpasses state-of-the-art topology-aware mesh generators in geometric fidelity and connectivity quality.

\end{abstract}

\section{Introduction}

Artist-created meshes exhibit compact, well-structured topology, characterized by adaptive vertex placement and clean, coherent face connectivity. Such structure is essential in production pipelines, where it supports reliable rigging and deformation for animation, high-quality shading, and efficient storage and rendering. Yet while topology-agnostic 3D generation has advanced rapidly, whether through VecSet representations~\citep{zhang20233dshape2vecset, zhang2024clay, zhao2025hunyuan3d, lai2026lattice} or structured latents~\citep{xiang2025structured, wu2026direct3d, li2026sparc3d}, these pipelines produce geometry as neural fields and extract surfaces via iso-surfacing, yielding dense, irregular tessellations devoid of artist-like structure. Directly generating meshes with such structure remains difficult because meshes are inherently hybrid representations: vertex coordinates are coupled with discrete, combinatorial connectivity of variable cardinality, not naturally captured by conventional generative models formulated over fixed-dimensional continuous spaces. To confront this discreteness, a prominent line of work tokenizes meshes into sequences and trains autoregressive models to generate faces token by token, often conditioned on point clouds~\citep{siddiqui2024meshgpt, chen2024meshxl, chen2025meshanythingv2, hao2024meshtron, weng2025scaling}. By modeling connectivity through sequential token prediction, these approaches produce structurally coherent, production-oriented meshes. However, serializing large, detailed meshes yields prohibitively long token streams, which inflates training cost, slows inference, and caps the attainable face count.

Flow-based topology-aware mesh generation has recently emerged as a compelling alternative. By modeling mesh elements in parallel within a continuous space rather than through sequential prediction, flow-based approaches offer substantially faster inference, admit stronger adherence to global conditions such as input images, and represent vertices and topology explicitly. Early attempts, including PolyDiff~\citep{alliegro2023polydiff} and MeshCraft~\citep{he2025meshcraft}, apply diffusion to per-face features, demonstrating the feasibility of continuous mesh generation but scaling poorly with mesh complexity. MeshFlow~\citep{li2026meshflow} instead adopts an edge-centric view, constructing a continuous topology space by concatenating vertex-pair features with dynamic padding over varying vertex counts and degrees. LATO~\citep{zhao2026lato} introduces a continuous topology representation via a vertex displacement field, compresses it into a structured latent space, and learns its distribution with flow matching; notably, vertices and connectivity are jointly encoded in a single unified latent, from which meshes are reconstructed through vertex subdivision, pruning, and a dedicated connectivity-prediction head.

Nevertheless, encoding vertices and connectivity in a single shared latent has a fundamental drawback: one flow model must jointly capture two statistically dissimilar signals, spatial positions and discrete combinatorial structure. This entanglement complicates flow learning and manifests as the broken surfaces observed in prior work. Our key observation is that the discreteness of connectivity is an artifact of modeling it \emph{jointly} with geometry. Once factorized, each factor admits a natural continuous representation: vertex positions are inherently continuous quantities, and, conditioned on realized vertices, connectivity reduces to relations among known points in space, expressible as continuous per-vertex features decoded into pairwise edge probabilities.

Building on this observation, we propose \methodname{}, a factorized explicit mesh generation pipeline that models vertices and connectivity with separate flow matching stages over two structured latent spaces. For vertices, we adopt the VAE architecture of LATO~\citep{zhao2026lato} to compress vertex positions alone, and additionally predict a per-vertex drift vector that refines each vertex from its voxel center at the finest resolution, compensating for quantization error. For connectivity, given the realized vertices, a topology latent is constructed as per-vertex features, sampled by a vertex-conditioned flow model and decoded into pairwise edge probabilities. Under this factorization, each flow models a simpler, homogeneous distribution, easing learning and alleviating the broken surfaces of shared-latent designs. The factorization further makes topology adaptive to vertex-level intervention, which we demonstrate in two settings: (i) part-wise generation, in which the voxel scaffold is partitioned and each part synthesized at full latent capacity, yielding substantially higher-resolution meshes than a monolithic latent permits; and (ii) topology-adaptive editing, in which manipulating first-stage vertices induces the corresponding connectivity without re-optimization. Notably, Nexus~\citep{wang2026nexus}, a concurrent and independent work, likewise generates vertices before connectivity, yet differs in both stages. For vertices, Nexus employs a coarse-to-fine octree diffusion model that generates level by level, whereas \methodname{} compresses vertices into a structured latent decoded in a single pass. For topology, Nexus formulates connectivity learning as graph autoencoding supervised by a SpaceTime loss, whereas \methodname{} trains the topology latent with a contrastive objective and reconstructs faces via loop detection. Extensive experiments show that \methodname{} surpasses state-of-the-art topology-aware mesh generators in geometric fidelity and connectivity quality, while uniquely supporting part-wise high-resolution generation and topology-adaptive editing.

Our contributions are summarized as follows:
\begin{itemize}
    \item We propose \methodname{}, a factorized flow matching framework that decomposes explicit mesh generation into a vertex flow and a connectivity flow over structured latents anchored to a shared voxel scaffold,.
    \item We introduce a per-vertex drift vector that compensates finest-resolution quantization error in vertex decoding, and a per-vertex topology latent can be decoded into pairwise edge probabilities.
    \item \methodname{} surpasses state-of-the-art topology-aware mesh generators while enabling vertex induced part-wise high-resolution generation and topology-adaptive editing.
\end{itemize}

\section{RELATED WORK}
\label{related_works}

\noindent \textbf{3D Shape Generation.}
Early 3D generation methods often relied on 2D diffusion priors due to the scarcity of large-scale 3D data. Representative approaches either optimize 3D representations through Score Distillation Sampling (SDS)~\citep{poole2022dreamfusion, chen2023fantasia3d, lin2023magic3d, tang2024dreamgaussian, yi2024gaussiandreamer}, or synthesize multi-view images before reconstructing 3D geometry~\citep{liu2023zero, liu2024syncdreamer, long2024wonder3d, shi2023zero123++, voleti2024sv3d, shi2024mvdream, wang2023imagedream}. Feed-forward reconstructors such as LRM~\citep{hong2024lrm} and subsequent variants improve efficiency and quality with multi-view supervision and alternative representations such as NeRFs and 3D Gaussians~\citep{li2024instant3d, kerbl20233d}. In parallel, 3D-native latent models generate point clouds or compact shape latents directly~\citep{gao2022get3d, liu2023meshdiffusion, nichol2022point, zhang20233dshape2vecset, zhang2024clay}, and recent structured latent methods further adopt sparse or high-resolution voxel representations~\citep{xiang2025structured, wu2026direct3d, li2026sparc3d, xiang2026native}. Despite their strong geometric modeling ability, these methods typically represent shapes as implicit fields, volumetric grids, Gaussians, or other non-mesh structures, and obtain meshes only through post-hoc surface extraction such as Marching Cubes~\citep{lorensen1998marching}. As a result, the extracted meshes are often dense and irregular, and their connectivity is not modeled as a first-class generative target.

\noindent \textbf{Autoregressive Mesh Generation.}
A line of work directly generates meshes by modeling vertices, edges, or faces as discrete sequences. PolyGen~\citep{nash2020polygen} and MeshGPT~\citep{siddiqui2024meshgpt} established the auto-regressive mesh generation paradigm, while later methods improve architecture design, conditioning, and tokenization~\citep{chen2024meshxl, chen2025meshanything, wang2024llama, weng2024pivotmesh}. Since naive face-wise encoding can require up to 9 times face number coordinate tokens, many recent methods focus on more compact mesh serialization, including adjacent mesh tokenization~\citep{chen2025meshanythingv2}, EdgeBreaker-style traversal~\citep{tang2025edgerunner, rossignac1999edgebreaker}, patchified block-wise indexing~\citep{weng2025scaling}, hierarchical BFS edge tokenization~\citep{song2025mesh}, tree-based sequencing~\citep{lionar2025treemeshgpt}, and frontier-aware expansion~\citep{lin2026meshripple}. Other works reduce the sequential burden with hierarchical Transformers~\citep{hao2024meshtron, nawrot2022hierarchical}, deterministic face prediction from generated vertices~\citep{kim2026fastmesh}, reinforced fine-tuning~\citep{zhao2025deepmesh, liu2026mesh}, or quadrilateral mesh generation~\citep{liu2025quadgpt}. These methods make mesh topology explicit, but they remain constrained by long token sequences, coordinate quantization, ordering sensitivity, and the difficulty of maintaining long-range topological consistency.

\noindent \textbf{Flow-based Explicit Mesh Generation.}
Continuous and non-autoregressive methods provide an alternative to sequential mesh token generation. DMesh~\citep{son2024dmesh} and DMesh++~\citep{son2025dmesh++} optimize differentiable face-existence probabilities, but rely on Delaunay-based candidate complexes. PolyDiff~\citep{alliegro2023polydiff} performs discrete denoising diffusion over quantized triangle soups, MeshCraft~\citep{he2025meshcraft} introduces continuous face-level tokens with a flow-based Diffusion Transformer, and SpaceMesh~\citep{shen2024spacemesh} represents topology with continuous vertex embeddings while still requiring connectivity recovery. More recent latent diffusion or flow methods, such as MeshFlow~\citep{li2026meshflow} and LATO~\citep{zhao2026lato}, move explicit mesh generation into compact latent spaces and generate geometry and connectivity more efficiently than long auto-regressive sequences. However, these methods still tend to couple vertex geometry and mesh connectivity within a shared representation or generate them in parallel. In contrast, our method explicitly factorizes mesh generation into high-resolution vertex generation and vertex-conditioned topology generation. This separation matches the different structures of the two problems: vertex geometry is decoded through local sparse refinement, while topology is modeled as a relational latent distribution conditioned on the generated vertex set.

\section{Method}
\begin{figure}[t]
    \begin{center}
    \end{center}
    \centering
    \includegraphics[width=\textwidth]{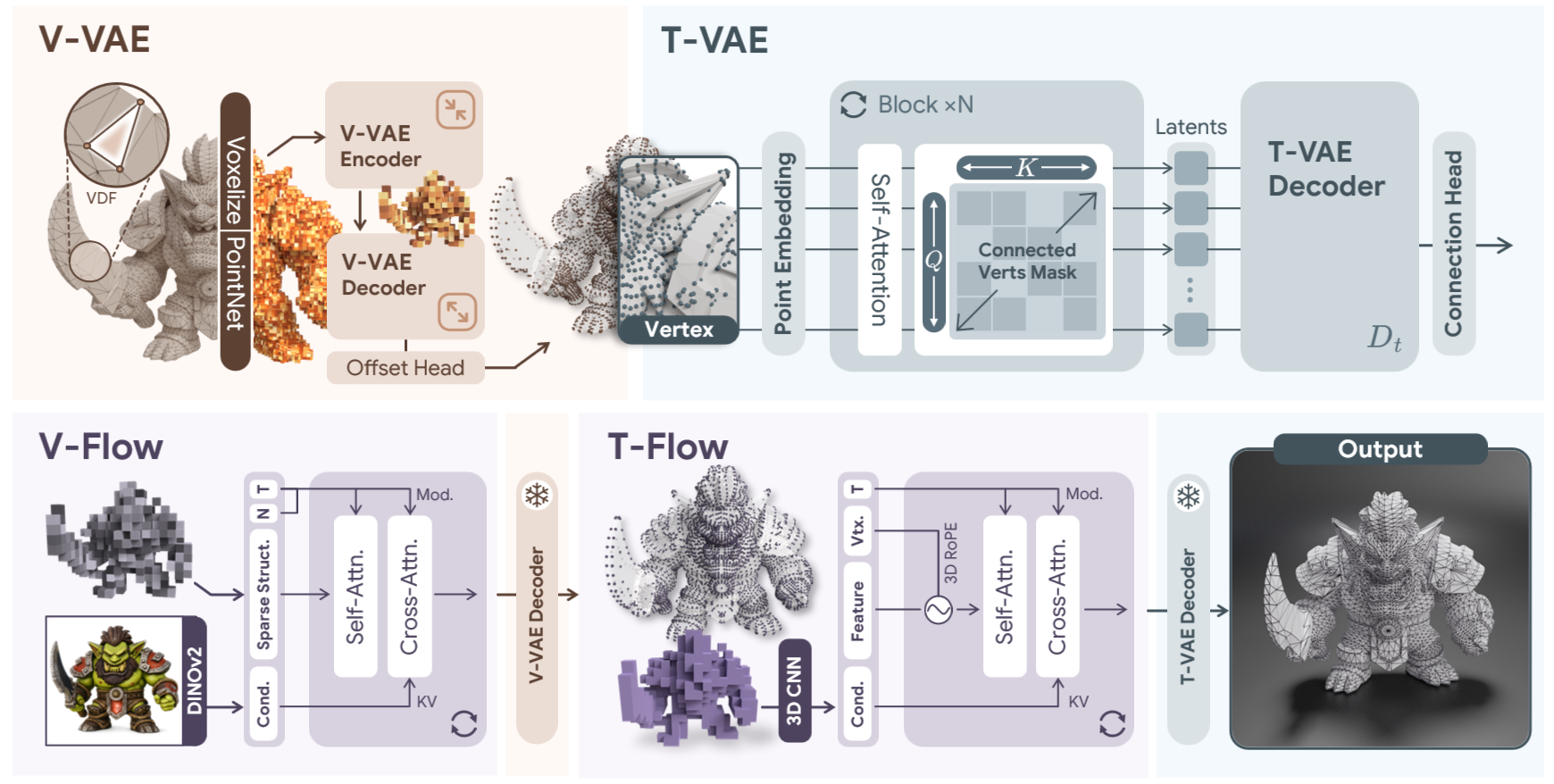}
    \caption{
    \textbf{Overview of \methodname{} pipeline.}
    \methodname{} factorizes explicit mesh generation into vertex and topology synthesis. V-VAE learns sparse vertex latents and reconstructs high-resolution vertex positions with sub-voxel offsets, while T-VAE learns topology latents through vertex-aware attention. At inference, V-Flow generates vertices conditioned on images with controllable vertex counts, and T-Flow predicts connectivity conditioned on the generated vertices.
    }
    \label{fig:pipeline}
\end{figure}

We present \methodname{}, a factorized sparse latent framework for explicit mesh generation, illustrated in Figure~\ref{fig:pipeline}. The framework comprises two  representation stages and two generative stages. The Vertex VAE (V-VAE) takes Vertex Displacement Field (VDF) as input~\citep{zhao2026lato}, compress VDF into a sparse structured latent; and decode into vertex voxels through iterative subdivision and pruning. To compensate for quantization at the finest voxel resolution, the decoder additionally predicts a per-vertex drift vector that refines each vertex away from its voxel center. Given realized vertices, the Topology VAE (T-VAE, \S\ref{sec:vae}) encodes connectivity as per-vertex features and recovers pairwise edge probabilities, from which faces are reconstructed via loop detection. Over these two latent spaces, we train a vertex flow and a vertex-conditioned connectivity flow (\S\ref{sec:flow}). 


\subsection{Vertex and Topology VAE}
\label{sec:vae}
Given a mesh $\mathcal{M}=(\mathbf{V},\mathbf{E})$, where
$\mathbf{V} = \{ \mathbf{v}_i | i=1\dots \mathrm{N} \} \subset \mathbb{R}^3$ denotes the vertice positions and
$\mathbf{E} = \{ \mathbf{e}_{ij} | i,j = 1\dots \mathrm{N} \}$ denotes the edge set.
We represent connectivity as a symmetric adjacency matrix
$\mathbf{A}=\{\mathbf{e}_{ij}\}\in\{0,1\}^{N\times N}$, where
\begin{equation}
\mathbf{e}_{ij}=
\begin{cases}
1, & \text{if } \mathbf{v}_i \text{ and } \mathbf{v}_j \text{ are connected},\\
0, & \text{otherwise}.
\end{cases}
\end{equation}
Since mesh connectivity is undirected, we have
$\mathbf{e}_{ij}=\mathbf{e}_{ji}$ and $\mathbf{e}_{ii}=0$.

Instead of encoding geometry and connectivity in a single shared latent space, we observe that the two signals admit fundamentally different decoding regimes. We train two autoencoders, i.e., the Vertex VAE (V-VAE) and Topology VAE (T-VAE), to define the latent spaces used by the flow matching models. We denote the encoder and decoder separately as: $\text{V-VAE} = ( \mathcal{E}_\mathbf{v}, \mathcal{D}_\mathbf{v} )$ and $\text{T-VAE} = ( \mathcal{E}_\mathbf{t}, \mathcal{D}_\mathbf{t} )$.

\noindent\textbf{Vertex VAE.}
The V-VAE compresses vertex positions into a structured latent and decodes
them back. Regressing vertex coordinates directly is ill-posed: the vertex
set is unordered and of variable cardinality. Following LATO~\citep{zhao2026lato},
we instead represent vertices through a Vertex Displacement Field (VDF), a
dense, continuous encoding anchored to the surface. Concretely, we sample a
point cloud $\mathcal{P}=\{\mathbf{p}_k\}_{k=1}^{K}$ on the mesh surface and
associate each point with its position $\mathbf{p}_k$, surface normal
$\mathbf{n}_k$, and displacement vector $\mathbf{d}_k$ pointing to a vertex of
the face on which it lies.

Each point is embedded with PointNet~\citep{qi2017pointnet}, and the point
features are aggregated into a sparse voxel grid. Rather than aggregating
directly at the latent resolution, we rasterize at high resolution
($1024^3$ in our implementation) and downsample through sparse 3D convolution
and attention to obtain the latent $\mathbf{z}_\mathbf{v}$. The decoder
$\mathcal{D}_\mathbf{v}$ reconstructs vertices through coarse-to-fine sparse refinement:
starting from $\mathbf{z}_\mathbf{v}$, it progressively subdivides voxels and
prunes those containing no vertex, and after each pruning step the surviving
voxel features are refined by sparse cross-attention with
$\mathbf{z}_\mathbf{v}$ to inject global context. At the finest level, the
decoder predicts for each surviving voxel $\hat{\mathbf{v}}_i$ a local offset
$\boldsymbol{\delta}_i$ from the voxel center to the enclosed vertex,
compensating for quantization error:
\begin{equation}
    \mathbf{z}_\mathbf{v} = \mathcal{E}_\mathbf{v}(\mathcal{P}), \qquad
    \{ \hat{\mathbf{v}}_i, \boldsymbol{\delta}_i \} = \mathcal{D}_\mathbf{v}(\mathbf{z}_\mathbf{v}),
    \label{eq:vertex_offset}
\end{equation}
with the reconstructed vertex positions given by
$\{ \hat{\mathbf{v}}_i + \boldsymbol{\delta}_i \}$.

We train the V-VAE with a pruning loss at each subdivision level $r$,
instantiated as an asymmetric focal loss $\mathcal{L}_{\mathrm{asy}}$~\citep{ridnik2021asymmetric} to counter the severe class imbalance between vertex-bearing and empty voxels; an offset regression loss $\mathcal{L}_{\mathrm{off}}$ (MSE) on $\boldsymbol{\delta}_i$; and a KL term $\mathcal{L}_{\mathrm{KL}}$ regularizing the latent. The full objective is
\begin{equation}
    \mathcal{L}_{v}
    =
    \sum_{r} \mathcal{L}^{r}_{\mathrm{asy}}
    +
    \mathcal{L}_{\mathrm{off}}
    +
    \mathcal{L}_{\mathrm{KL}},
    \label{eq:vertex_loss}
\end{equation}
where loss weights are omitted for clarity.

\paragraph{Topology VAE.}
The T-VAE learns a latent representation $\mathbf{z}_\mathbf{t}$ of mesh
connectivity from which the adjacency matrix $\mathbf{A}$ can be
reconstructed. We cast this as per-vertex feature learning over the realized
vertex set $\hat{\mathbf{V}} = \{ \hat{\mathbf{v}}_i \}_{i=1}^{N}$: the
latent is a set of per-vertex features aligned with $\hat{\mathbf{V}}$, so
that connectivity information is stored at the vertices it involves.

The encoder $\mathcal{E}_\mathbf{t}$ embeds vertex positions and processes the
resulting tokens with a transformer:
\begin{equation}
    \mathbf{z}_\mathbf{t} = \mathcal{E}_\mathbf{t}(\hat{\mathbf{V}})
    = \mathrm{SelfAttn}^{8}\!\left(\mathrm{PosEmb}(\hat{\mathbf{V}})\right),
    \label{eq:topo_encoder}
\end{equation}
where $\mathrm{PosEmb}(\cdot)$ denotes positional encoding of vertex coordinates. Crucially, ground-truth connectivity enters the encoder only through an attention mask: in every layer, a vertex token may attend solely to itself and to vertices adjacent to it in the mesh,
\begin{equation}
    M_{ij}=
    \begin{cases}
    0, & \mathbf{e}_{ij}=1 \ \text{or}\ i=j,\\
    -\infty, & \text{otherwise},
    \end{cases}
    \qquad
    \mathrm{Attn}(i,j)=
    \mathrm{softmax}_{j}\!\left(
    \frac{q_i^\top k_j}{\sqrt{d}}+M_{ij}
    \right),
    \label{eq:edge_mask}
\end{equation}
where $\mathbf{e}_{ij}\in\{0,1\}$ is the ground-truth adjacency. Since the
mask is the sole channel through which edges are observed, the encoder is
forced to absorb connectivity into the per-vertex latent.

The decoder $\mathcal{D}_\mathbf{t}$ recovers $\mathbf{A}$ from $\mathbf{z}_\mathbf{t}$
using a vanilla transformer with unmasked self-attention.
Denoting by $\mathbf{h}_i$ as the final-layer hidden state of vertex $i$, an edge
probability is predicted for a vertex pair by an MLP classifier following
LATO~\citep{zhao2026lato}:
\begin{equation}
    \mathbf{h} = \mathcal{D}_\mathbf{t}(\mathbf{z}_\mathbf{t}), \qquad
    \hat{\mathbf{e}}_{ij} =
    \sigma  \big [
    \text{MLP}(\mathbf{h}_i \oplus \mathbf{h}_j) +
    \text{MLP}(\mathbf{h}_j \oplus \mathbf{h}_i)
     \big ] ,
    \label{eq:edge_decoder}
\end{equation}
where $\oplus$ denotes concatenation; evaluating both orders and averaging the
logits enforces symmetry of the predicted adjacency. Since scoring all $N^2$
pairs is intractable for large meshes, training evaluates a candidate set
$\mathcal{C}$ comprising each vertex's true connected vertices as positives, top-$K$ nearest neighbors, which
concentrate and hard negatives, supplemented with uniformly sampled random pairs as easy negatives.

The T-VAE is trained with an asymmetric focal loss~\citep{ridnik2021asymmetric} over
$\mathcal{C}$ and a KL regularization on the topology posterior:
\begin{equation}
    \mathcal{L}_{t}
    =
    \frac{1}{|\mathcal{C}|}
    \sum_{(i,j)\in\mathcal{C}}
    \mathcal{L}_{\mathrm{asy}}
    \left(
    \hat{\mathbf{e}}_{ij},\mathbf{e}_{ij}
    \right)
    +
    \mathcal{L}_{\mathrm{KL}}.
    \label{eq:topo_loss}
\end{equation}
At inference, faces are recovered from the predicted edge set via loop detection.

\subsection{Factorized Mesh Generation}
\label{sec:flow}
With both autoencoders trained, we learn flow models over their latent spaces. Given a task conditio, such as an image or a point cloud, \methodname{} generates a mesh in three steps: it first produces a coarse sparse structure, then samples the vertex latent on its active voxels and
decodes high-resolution vertices, and finally, conditioned on the realized vertices, samples the topology latent and decodes pairwise connectivity.

\paragraph{Vertex Flow Model.}
Following TRELLIS~\citep{xiang2025structured}, we train a flow model to generate a coarse sparse structure $\hat{\mathcal{S}}$ from the task prompt, an image or a point cloud, which specifies the active spatial support for vertex latent tokens. A sparse flow transformer, the V-Flow model, then samples the vertex latent $\mathbf{z}_\mathbf{v}$ on the active cells of $\hat{\mathcal{S}}$, and the vertex decoder $\mathcal{D}_\mathbf{v}$ reconstructs the vertex set via progressive sparse refinement. 

The V-Flow model is conditioned on visual features $\mathbf{c}_\text{img}$, DINO~\citep{caron2021emerging} features extracted from a random rendering of the training mesh, injected via cross-attention. To make the vertex count controllable, we further condition the flow on a count factor  $\mathbf{c}_\text{vn} = \log_2 N$, the logarithm of the target vertex number, embedded by an MLP and injected via adaLN together with the timestep embedding. During training, $N$ is the ground-truth count; at inference it is user-specified, providing direct control over the generated vertex number. The vertex flow is trained with the rectified-flow objective
\begin{equation}
    \mathcal{L}_{\text{V-Flow}}
    =
    \mathbb{E}_{\mathbf{z}_\mathbf{v},\,\tau,\,\epsilon,\,\mathbf{c}_\text{img},\,\mathbf{c}_\text{vn}}
    \left[
    \left\|
    v_{\theta}(\mathbf{z}_\mathbf{v}^\tau, \tau, \mathbf{c}_\text{img}, \mathbf{c}_\text{vn})
    -
    (\epsilon - \mathbf{z}_\mathbf{v})
    \right\|_2^2
    \right],
    \label{eq:vflow_loss}
\end{equation}
where $\epsilon \sim \mathcal{N}(0, I)$ and
$\mathbf{z}_\mathbf{v}^\tau = (1-\tau)\,\mathbf{z}_\mathbf{v} + \tau\epsilon$
for $\tau \in [0,1]$.

\paragraph{Topology Flow Model.}
Given the realized vertices $\hat{\mathbf{V}}$ decoded from the sampled vertex latent, a topology flow transformer samples the topology latent
$\mathbf{z}_\mathbf{t}$ over per-vertex tokens. Vertex positions enter the flow directly through the token construction: each token is associated with a vertex in $\hat{\mathbf{V}}$ and carries its position via 3D positional encoding, so the sampled latent is conditioned on the realized vertices by
design. We additionally provide a coarse geometric context $\mathbf{c}_\text{g} = \text{3DCNN}(\hat{\mathcal{S}})$, which supplies global shape information that guides connectivity prediction beyond local vertex neighborhoods.To support unconditional sampling without coarse geometric context, e.g., when editing existing meshes, we randomly drop $\mathbf{c}_\text{g}$ during training. We further apply RMS normalization to queries and keys in the attention blocks~\citep{zhang2019root}, stabilizing attention over vertex sets of widely varying size. The topology flow is trained with the same objective:
\begin{equation}
    \mathcal{L}_{\text{T-Flow}}
    =
    \mathbb{E}_{\mathbf{z}_\mathbf{t},\,\tau,\,\epsilon,\,\mathbf{c}_\text{g}}
    \left[
    \left\|
    v_{\phi}(\mathbf{z}_\mathbf{t}^\tau, \tau, \mathbf{c}_\text{g})
    -
    (\epsilon - \mathbf{z}_\mathbf{t})
    \right\|_2^2
    \right],
    \label{eq:tflow_loss}
\end{equation}
with $\mathbf{z}_\mathbf{t}^\tau$ defined analogously.

At inference, the task condition first drives generation of the sparse structure $\hat{\mathcal{S}}$. The vertex flow then samples the vertex latent
under the specified count condition $\mathbf{c}_\text{vn}$, and $\mathcal{D}_\mathbf{v}$ decodes the high-resolution vertex set. Conditioned on these vertices, the topology flow samples the topology latent, and $\mathcal{D}_\mathbf{t}$ predicts pairwise connectivity; by default we score
all vertex pairs, an $O(N^2)$ operation that remains tractable since each pair requires only a lightweight MLP evaluation over precomputed per-vertex
features, and faces are finally recovered via loop detection.

\subsection{Vertex-Induced Adaptive Topology}
The factorized design carries a practical consequence beyond simple generation: since the topology flow is conditioned on realized vertices, \emph{any} modification to the vertex set, whether user edits or programmatic recomposition, can be propagated to connectivity simply by re-running the T-Flow. We demonstrate two applications of this property: topology-adaptive mesh editing and part-wise refined generation.

\begin{figure}[t]
    \begin{center}
    \end{center}
    \centering
    \includegraphics[width=\textwidth]{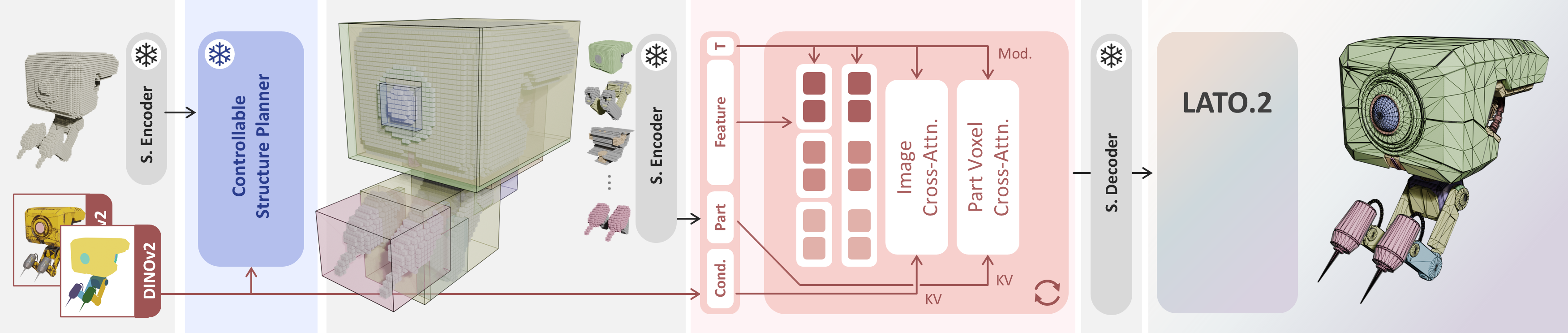}
    \caption{\textbf{Part-wise generation pipeline.} A controllable structure planner~\citep{yang2025omnipart} augments the sparse structure generated from the image or point cloud condition with part-level bounding boxes. The resulting box-aware structure is encoded into part features and fused with image features via cross-attention between image tokens and part voxels. The structure decoder then predicts a part-aware sparse structure, from which \methodname{} generates vertices for each part. The T-Flow finally predicts connectivity either jointly over the union of all part vertices, or per part with the results composed via stitching.}
    \label{fig:multi_part_pipeline}
\end{figure}

\paragraph{Topology-Adaptive Mesh Editing.}
In practical mesh editing scenarios, vertex deformation and topology adjustment are inherently coupled: while modifying vertex positions is straightforward, recovering valid connectivity after such changes remains labor-intensive and often requires specialized heuristics. Our factorized representation decouples these two processes, allowing users to freely manipulate vertices while delegating topology reconstruction to the T-Flow. Given an edited vertex set, obtained either from the V-Flow output or an existing mesh, we re-sample the topology latent conditioned on the edited set and decode the adapted connectivity. This enables topology-aware editing operations without explicitly designing topology manipulation rules.

\begin{figure}[h]
    \centering
    \includegraphics[width=\textwidth]{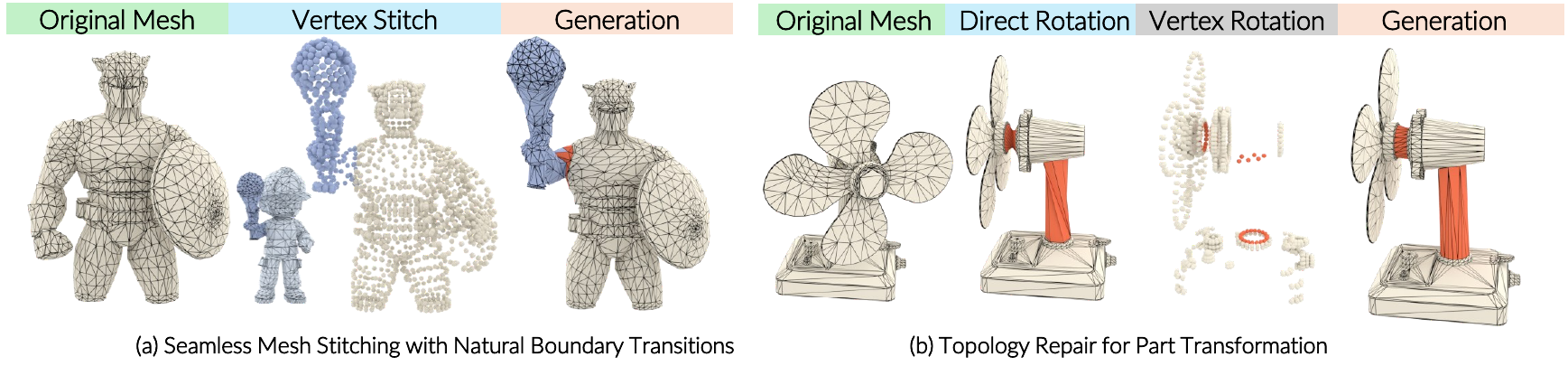}
    \caption{
    \textbf{Topology adaptive mesh editing.} a) Mesh stitching: given the union of vertex sets cropped from different meshes, our T-Flow generates seamless connectivity across the junctions; newly synthesized faces are highlighted in \textcolor{red}{red}. b) Part transformation: naively rotating a part while keeping the original connectivity fixed causes face stretching and self-intersections (artifacts highlighted in \textcolor{red}{red}, left), whereas regenerating connectivity with our T-Flow yields an intersection-free, topologically coherent mesh (\textcolor{red}{red}, right).
    }
    \label{fig:app}
\end{figure}

\noindent\textit{Mesh stitching.} To combine multiple meshes or mesh parts, we first transform them into a shared coordinate frame through alignment and rescaling, and then take the union of their vertex sets. The T-Flow subsequently generates connectivity over the combined vertices, automatically producing coherent topology across the newly formed interfaces. As shown in Figure~\ref{fig:app}(a), the generated topology effectively bridges the junction regions while preserving the geometric structures of individual components.

\noindent\textit{Part transformation.} For local deformation, a subset of vertices can be independently transformed through rotation or translation. Since the internal topology of both transformed and unchanged regions remains valid, only the transition area, where faces connect moved and unmoved vertices, requires topology adaptation. We identify these affected regions and regenerate their connectivity using the T-Flow conditioned on the updated vertex positions. Figure~\ref{fig:app}(b) illustrates that the originally invalidated connectivity caused by part transformation can be automatically repaired through our topology generation process.

\begin{figure}[h]
    \centering
    \includegraphics[width=\textwidth]{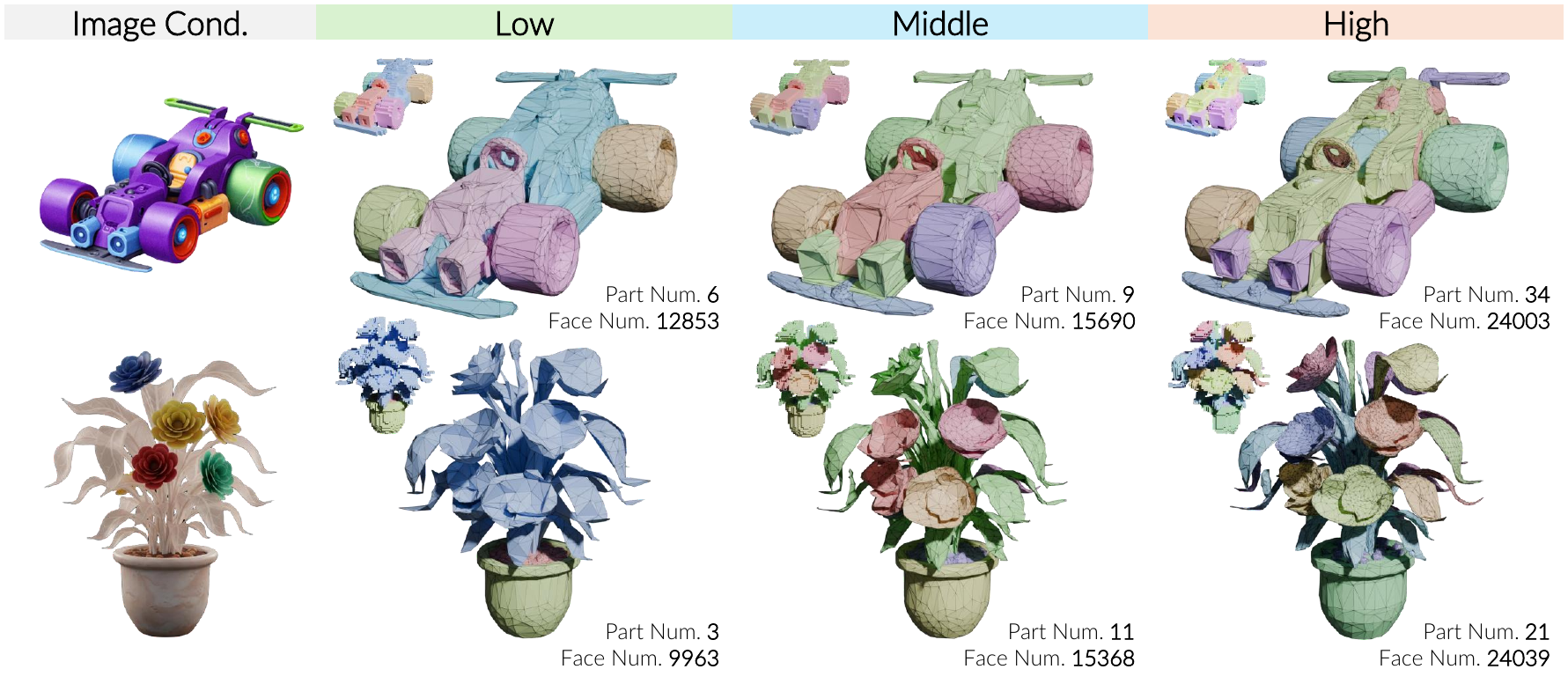}
    \caption{
    \textbf{Part-wise generation and refinement.} The coarse structure is partitioned into parts, each rescaled to occupy the full latent volume and generated at full latent capacity, yielding substantially denser and more detailed meshes. As the number of parts increases, the face count grows accordingly and finer geometric detail emerges.
    }
    \label{fig:app_part}
\end{figure}

\paragraph{Part-wise Generation and Refinement.}
A monolithic latent representation inherently limits mesh resolution, as the maximum number of decoded vertices is constrained by the latent grid capacity. Our factorized formulation overcomes this limitation by enabling independent generation and subsequent topology integration across multiple parts. Following a part decomposition strategy similar to OmniPart~\citep{yang2025omnipart}, we partition the coarse structure $\hat{S}$ into multiple parts and generate each part independently. Specifically, each part is normalized to occupy the full latent volume, allowing the V-Flow to sample vertices at the maximum latent resolution. The generated vertices are then transformed back into the original coordinate frame. Finally, the union of all part vertices is provided as input to one or multiple T-Flow passes to produce coherent connectivity. Figure~\ref{fig:multi_part_pipeline} illustrates the overall part-wise generation pipeline. This strategy enables the generation of meshes with substantially higher vertex density and finer geometric details compared with monolithic generation under the same latent resolution. Furthermore, the same mechanism naturally supports mesh refinement: a local region of an existing or generated mesh can be extracted, regenerated with increased vertex capacity, and reintegrated through the topology-adaptive stitching procedure described above. Figure~\ref{fig:app_part} presents representative examples of part-wise generation and refinement, demonstrating that increasing the number of generated parts progressively improves geometric resolution and mesh quality.

\begin{figure}[h]
    \begin{center}
    \end{center}
    \centering
    \includegraphics[width=\textwidth]{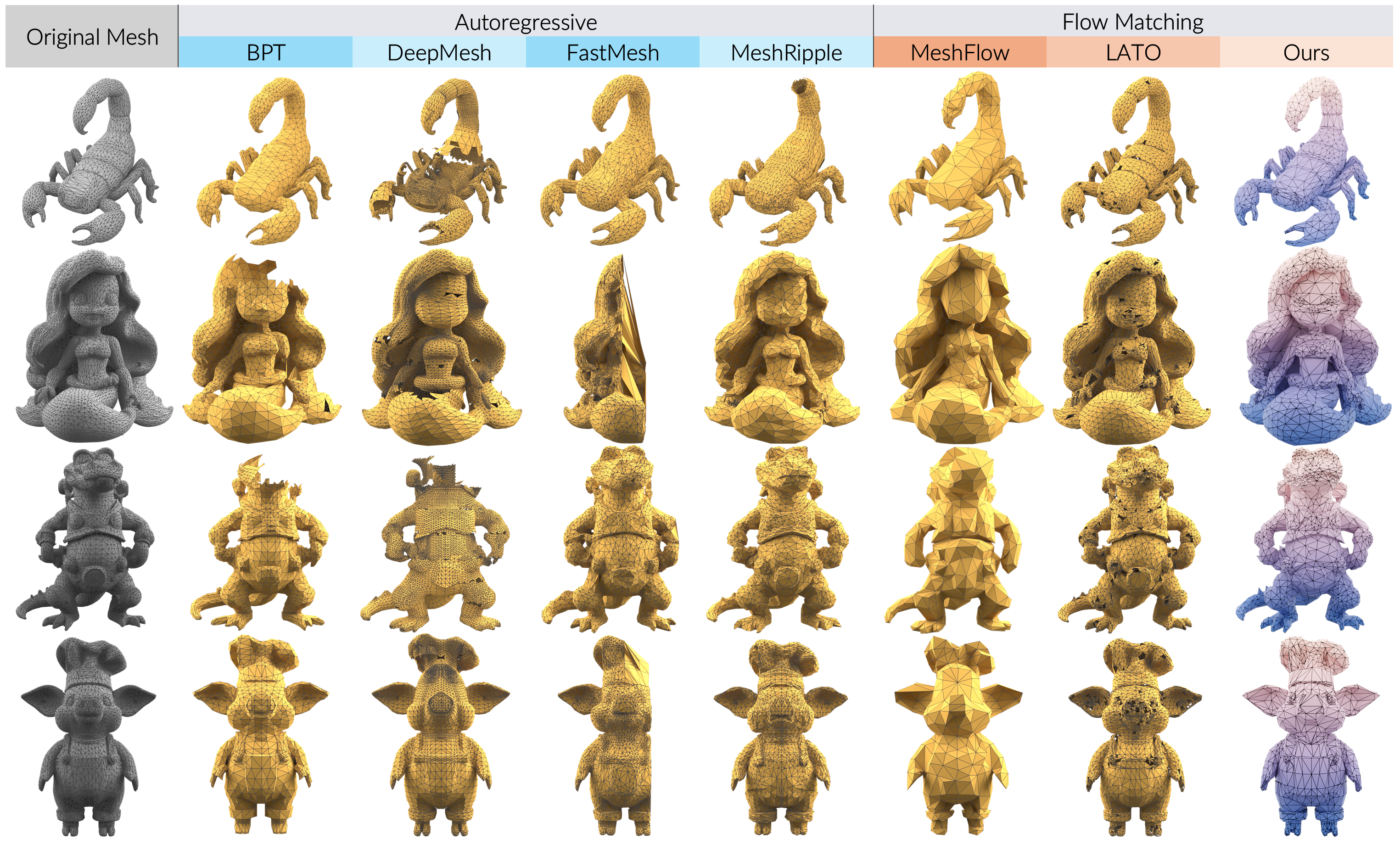}
    \caption{\textbf{Qualitative comparison on geometry-conditioned mesh generation.} Given point cloud or voxel conditions, baseline methods exhibit missing or broken surface regions, whereas our method produces complete shapes with more coherent topology.}
    \label{fig:quatitave_results}
\end{figure}

\section{EXPERIMENTS}
\label{experiments}
We evaluate \methodname{} from four perspectives: mesh reconstruction quality, conditional mesh generation quality, topology fidelity, and the effectiveness of the proposed factorized formulation. We compare against representative autoregressive mesh generators and recent latent-flow-based topology-aware mesh generation methods. Furthermore, we conduct extensive ablation studies to analyze the contributions of the vertex autoencoder, topology generation module, and vertex-count controllability.

\subsection{EXPERIMENTAL SETTINGS}

\textbf{Datasets.}
We train our model on a combination of datasets, including TRELLIS500K~\citep{xiang2025structured}, Objaverse~\citep{deitke2023objaverse}, Objaverse-XL~\citep{deitke2023objaversexl}, 3D-FUTURE~\citep{fu20213d}, Toys4K~\citep{stojanov2021using}, and ABO~\citep{collins2022abo}, comprising approximately 450K 3D assets in total. In addition, we construct a procedural dataset containing 100K synthetic meshes to further improve the diversity of geometric and topological structures. These synthetic meshes are generated from basic geometric primitives, including cubes, cylinders, cones, and spheres. We apply a series of geometric operations, such as subdivision, deformation, twisting, warping, and primitive composition, to create shapes with diverse vertex densities, local curvature distributions, and connectivity patterns.

\textbf{Implementation Details.}
For each mesh, we uniformly sample 819,200 surface points, where each point is associated with its position, normal, and vertex displacement field. The V-VAE first encodes point-wise features using PointNet and aggregates them into a $1024^3$ sparse voxel grid. The voxel features are subsequently compressed through sparse 3D convolutional downsampling and a Sparse Transformer with a hidden dimension of 512 and 8 attention heads, resulting in a $64^3$ sparse vertex latent representation with 32 channels. The V-VAE decoder reconstructs high-resolution vertices by progressively upsampling the latent representation to $1024^3$ through occupancy-pruned subdivision, latent cross-attention, and a sub-voxel offset regression head. V-Flow adopts a 12-block conditional Flow Matching Transformer architecture adapted from TRELLIS~\citep{xiang2025structured}, conditioned on DINOv2 image features and the target vertex count condition. For topology modeling, vertex coordinates are discretized into $K=1024$ bins and embedded using Fourier features with a width of $d=768$. The T-VAE employs adjacency-masked attention followed by full vertex attention, with a per-vertex latent dimension of $d_z=16$. It decodes mesh connectivity through a topology stack and a pairwise MLP classifier. The T-Flow generates topology latents conditioned on the vertex set, and further incorporates vertex positional encodings to capture spatial relationships among vertices during latent generation.
Counting trainable parameters only, the V-VAE, V-Flow, T-VAE, and T-Flow contain about 320M, 160M, 180M, and 240M parameters, respectively.
The VAE counts include both encoder and decoder, while frozen
conditioning encoders are excluded.
All models are optimized using AdamW with a learning rate of $10^{-4}$. Training is conducted on 8 NVIDIA H100 GPUs. The V-VAE and T-VAE are trained for 4 days and 1 day, respectively. The V-Flow is trained for 7 days with an effective batch size of 256, while the T-Flow is trained for 2 days using dynamic batching based on the vertex-token budget.

\begin{figure}[h]
    \begin{center}
    \end{center}
    \centering
    \includegraphics[width=\textwidth]{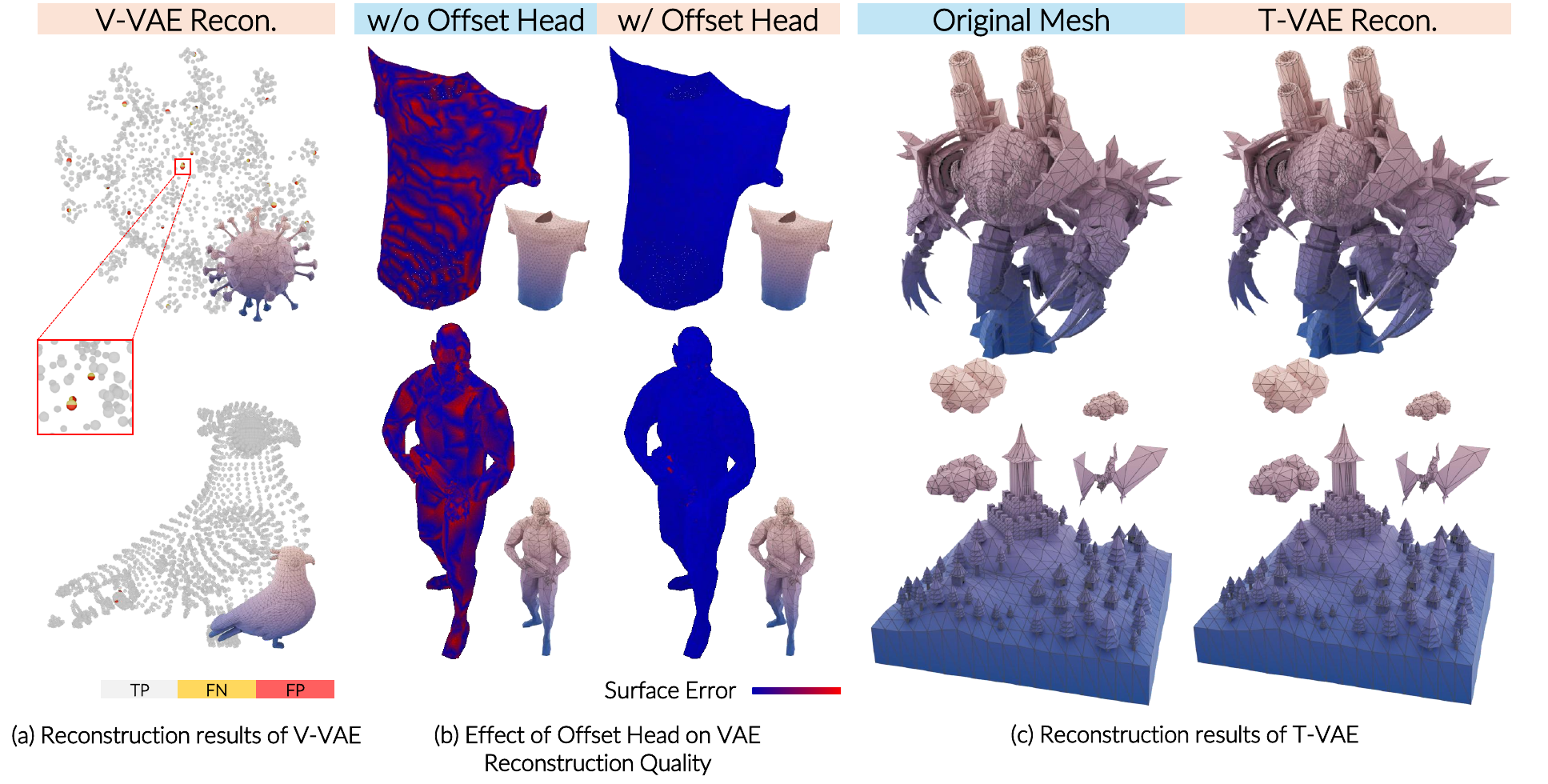}
    \caption{\textbf{Visualization on V-VAE and T-VAE reconstruction performances.} 
a) Visualization of V-VAE vertex reconstruction. Gray vertices denote correctly reconstructed vertices, yellow vertices denote missing ground-truth vertices, and red vertices denote false positive predictions. 
b) Effect of the offset head on VAE reconstruction quality, significant less error presented with offset head. 
c) Topology reconstruction by T-VAE. Given the original mesh, T-VAE reconstructs connectivity from the learned topology latent perfectly.}
    \label{fig:vae_perf}
\end{figure}

\begin{table}[h]
\centering
\caption{Quantitative comparison on V-VAE performance.}
\label{tab:vae_quantitative_comparison}
\begin{tabular}{ccccc}
\toprule
\textbf{Method} & \textbf{Quantization Levels} & \textbf{CD(L2)}$\downarrow$ & \textbf{CD(L1)}$\downarrow$ & \textbf{HD}$\downarrow$  \\
\midrule
MeshGPT & 128 & \underline{0.0013} & 0.0185 & 0.0955 \\
PivotMesh & 128 & 0.0074 & 0.0395 & 0.2227 \\
MeshCraft & 256 & 0.0106 & 0.0524 & 0.2842 \\
LATO & 512 & \textbf{0.0000} & \underline{0.0038} & \underline{0.0083} \\
\midrule
\textbf{Ours (V-VAE)} & \textbf{1024 + $\boldsymbol{\delta}$ (Float)} & \textbf{0.0000} & \textbf{0.0003} & \textbf{0.0069} \\
\bottomrule
\end{tabular}
\end{table}

\textbf{Baselines.}
We compare \methodname{} with two categories of explicit mesh generation approaches. The first category is autoregressive mesh generators, including MeshAnythingV2~\citep{chen2025meshanything}, BPT~\citep{weng2025scaling}, FastMesh~\citep{kim2026fastmesh}, DeepMesh~\citep{zhao2025deepmesh}, MeshSilkSong~\citep{song2025mesh}, and MeshRipple~\citep{lin2026meshripple}. These methods generate explicit meshes by autoregressively predicting mesh tokens or by factorizing mesh components into separately modeled representations. The second category consists of latent-flow-based topology-aware mesh generators, including LATO~\citep{zhao2026lato} and MeshFlow~\citep{li2026meshflow}. For a fair comparison, we evaluate all methods under identical input conditions and on the same test split whenever official checkpoints or publicly released generation results are available. For methods that generate meshes with different resolutions, we follow their default inference configurations and directly evaluate the resulting meshes without additional post-processing.

\textbf{Metrics.}
We evaluate both surface reconstruction fidelity and vertex reconstruction accuracy. For mesh-level reconstruction and conditional generation, we uniformly sample points from the surfaces of the predicted and reference meshes, and compute Chamfer Distance (CD), Hausdorff Distance (HD), and Normal Consistency (NC). CD measures the average bidirectional geometric discrepancy between the two surfaces, HD evaluates the maximum surface deviation, and NC quantifies the consistency of local surface orientations. For vertex-level reconstruction, we evaluate the predicted sparse occupancy representation at the finest resolution using precision, recall, F1 score, and Intersection-over-Union (IoU).

\begin{table}[h]
\centering
\caption{Quantitative comparison with topology-aware mesh generation methods. Our method achieves the strongest performance across all metrics.}
\label{tab:quantitative_comparison}
\begin{tabular}{lccccc}
\toprule
\textbf{Method} & \textbf{Type} & \textbf{CD(L2)}$\downarrow$ & \textbf{CD(L1)}$\downarrow$ & \textbf{HD}$\downarrow$ & $\boldsymbol{|\mathrm{NC}|}\uparrow$ \\
\midrule
MeshAnythingV2        & AR   & 0.1083 & 0.1505 & 0.2318 & 0.6946 \\
FastMesh              & AR   & 0.0822 & 0.1163 & 0.1397 & 0.6939 \\
MeshSilkSong          & AR   & 0.0654 & 0.0937 & 0.1455 & 0.7759 \\
BPT                   & AR   & 0.0603 & 0.0862 & 0.1067 & 0.8111 \\
DeepMesh              & AR   & 0.0529 & 0.0765 & 0.0946 & 0.8218 \\
MeshRipple            & AR   & 0.0458 & 0.0668 & 0.0941 & 0.8174 \\
\midrule
MeshFlow              & Flow &  0.0455      &  0.0668      &  0.0772      &  0.8227      \\
LATO                  & Flow & \underline{0.0421} & \underline{0.0617} & \underline{0.0738} & \underline{0.8262} \\
\midrule
\textbf{Ours} 
                      & Flow & \textbf{0.0407} & \textbf{0.0596} & \textbf{0.0657} & \textbf{0.8341} \\
\bottomrule
\end{tabular}
\end{table}

\subsection{Quantitative Analysis}

\textbf{V-VAE Performance.}
We first evaluate the performance of V-VAE. As reported in Tab.~\ref{tab:vae_quantitative_comparison}, \methodname{} achieves superior reconstruction performance compared with existing mesh autoencoding baselines. Compared with quantized token-based representations and the vertex autoencoder of LATO, our continuous sub-voxel offset prediction significantly reduces geometric reconstruction errors. This indicates that, after sparse voxel pruning, the dominant source of reconstruction error arises from discretization artifacts, which can be effectively mitigated by predicting continuous offsets within each voxel. These results validate the importance of sub-voxel offset modeling for high-fidelity vertex reconstruction.

\textbf{Geometry Conditioned Mesh Generation.}
We further evaluate the geometry conditioned mesh generation pipeline using point-cloud as conditions. Following previous protocols for shape-conditioned mesh generation, we conduct experiments on held-out assets and compare against both autoregressive mesh generators and latent-flow-based approaches. As shown in Tab.~\ref{tab:quantitative_comparison}, \methodname{} achieves the strongest overall performance across both geometric and topology-related metrics. Compared with autoregressive approaches, our sparse latent generation avoids the need for excessively long mesh token sequences, enabling more stable generation of meshes with higher vertex counts. Compared with existing latent-flow methods, our explicit factorization of vertex geometry and topology provides more accurate geometric reconstruction while maintaining more coherent connectivity.

%
%

\textbf{Topology Generation.}
To isolate the effect of topology modeling, we compare different combinations of vertex and topology sources in Tab.~\ref{tab:pair_quantitative_comparison}. Using ground-truth vertices with the topology autoencoder provides an upper bound for connectivity reconstruction. Replacing the topology autoencoder with the topology flow measures the generative error of topology sampling, while replacing ground-truth vertices with generated vertices evaluates the full pipeline. The results show that the topology flow preserves most of the reconstruction quality and that the factorized design localizes errors to the generative stages rather than the autoencoders themselves.

\begin{table}[t]
\centering
\caption{
\textbf{Evaluation of topology generation under various vertex inputs.} Each row pairs a vertex source with a topology module. GT-Verts denotes ground-truth vertices; GT-Verts + T-VAE reconstructs topology from the ground-truth mesh and serves as the performance upper bound. Rows 2-4 evaluate T-Flow sampling conditioned on ground-truth, V-VAE reconstructed, and V-Flow generated vertices, respectively, isolating the error introduced by each stage.}

\label{tab:pair_quantitative_comparison}
\begin{tabular}{cccccc}
\toprule
\textbf{Vtx.} & \textbf{Topo.} & \textbf{CD(L2)}$\downarrow$ & \textbf{CD(L1)}$\downarrow$ & \textbf{HD}$\downarrow$ & $\boldsymbol{|\mathrm{NC}|}\uparrow$ \\
\midrule
GT-Verts & T-VAE                  & \textbf{0.0351} & \textbf{0.0478} & \textbf{0.0859} & \textbf{0.8615} \\
GT-Verts & T-Flow                & 0.0399 & 0.0543 & 0.1047 & 0.8199 \\
V-VAE & T-Flow                   & \underline{0.0393} & \underline{0.0538} & \underline{0.0993} & \underline{0.8201} \\
V-Flow & T-Flow  & 0.0407 & 0.0570 & 0.1029 & 0.8051 \\
\bottomrule
\end{tabular}
\end{table}

\subsection{Qualitative Analysis}
\paragraph{Qualitative Comparison.}
Fig.~\ref{fig:quatitave_results} shows qualitative comparisons with autoregressive baselines and latent-flow topology-aware baselines. Autoregressive methods often produce broken surfaces, missing thin structures, or irregular local connectivity when the target mesh contains complex geometry. Latent-flow baselines generate more coherent global shapes, but may still lose fine geometric details or produce less faithful connectivity. In contrast, \methodname{} preserves fine structures while maintaining cleaner mesh connectivity, benefiting from the separation of high-resolution vertex generation and vertex-conditioned topology generation.

\begin{figure}[h]
    \begin{center}
    \end{center}
    \centering
    \includegraphics[width=\textwidth]{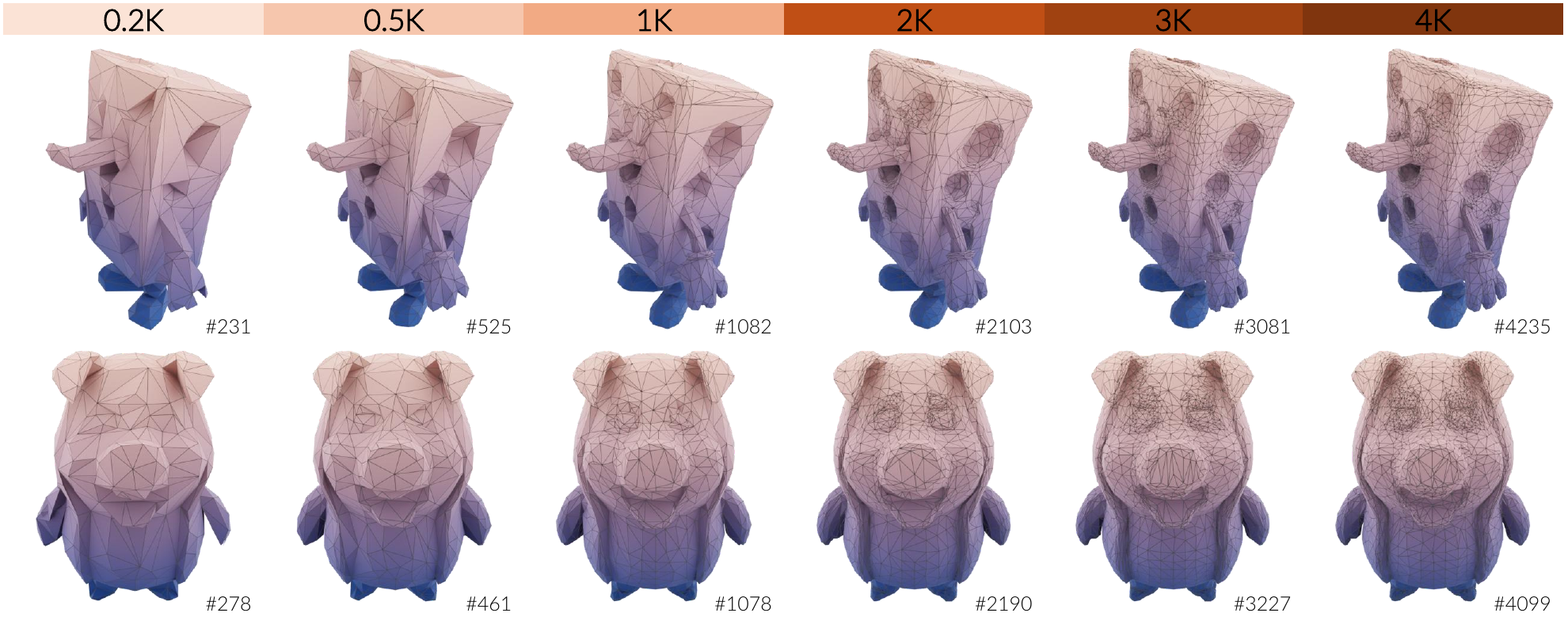}
    \caption{
    \textbf{Vertex number controlled generation.}
    The explicit vertex number condition of V-Flow enables controllable mesh resolution by specifying the target number of vertices. Given the same structure voxels, increasing the vertex budget from $0.2$K to $4$K produces progressively denser meshes with finer geometric details, while the topology flow automatically adapts connectivity to the generated vertex set and preserves the underlying shape structure.
    }
    \label{fig:density}
\end{figure}

\paragraph{Vertex Number Controllability.}
Fig.~\ref{fig:density} visualizes vertex-count controllability. Given the same input condition, increasing the target vertex budget produces progressively denser meshes while preserving the object-level geometry. This demonstrates that the explicit sparse vertex generation process can serve as a practical control mechanism for mesh resolution.

\begin{figure}[h]
    \centering
    \includegraphics[width=\textwidth]{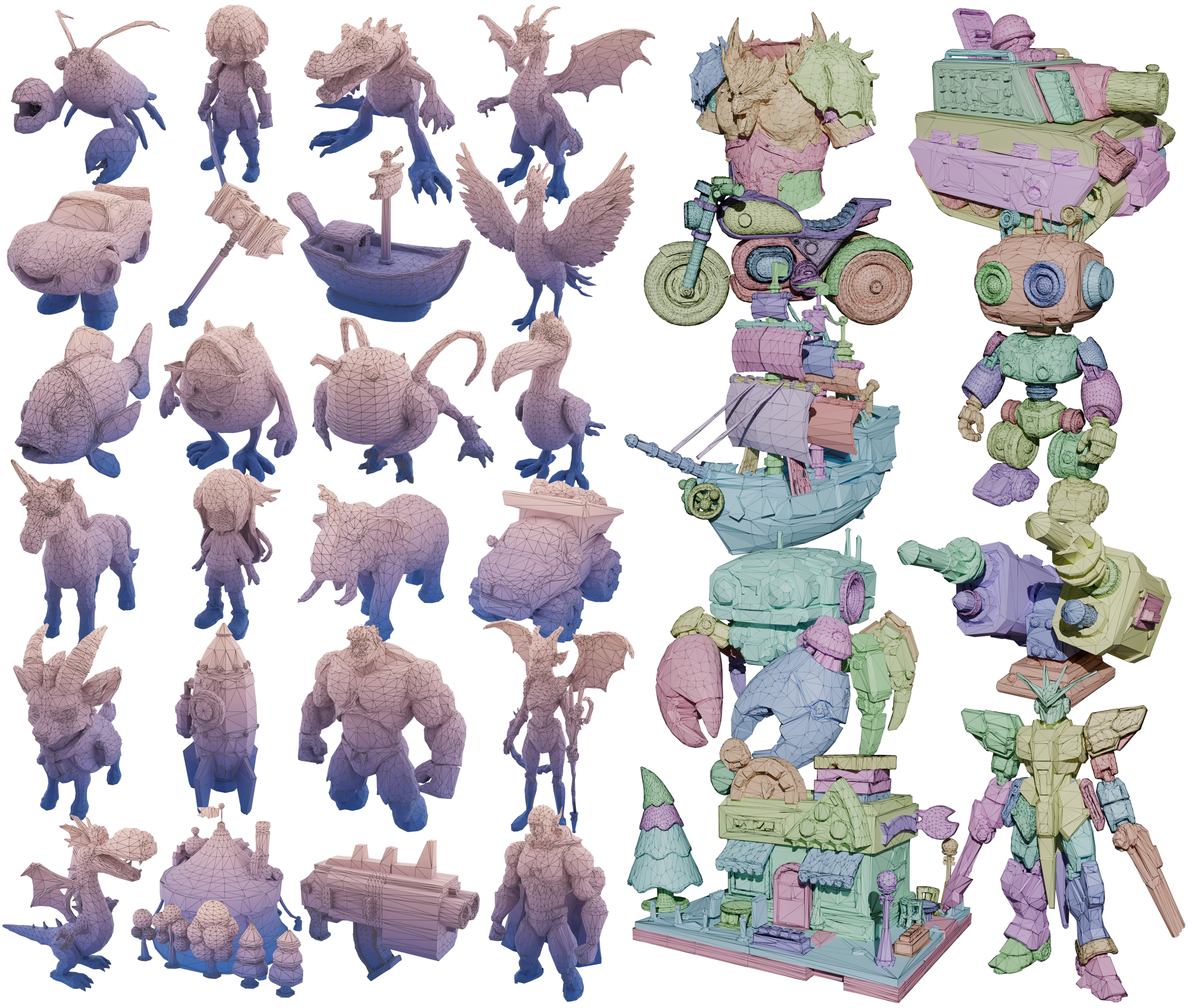}
    \caption{\textbf{Generation gallery.} Our method can generate various types of meshes with one pass generation. With multi-part generation, \methodname{} is capable of generating meshes with high face count and extreme fine details.}
    \label{fig:gallery}
\end{figure}

\textbf{Gallery.}
We further present a gallery of generated meshes obtained through both monolithic generation and the multi-part generation process in Fig.~\ref{fig:gallery}. The diverse examples, ranging from high-density geometric structures to stylized artistic shapes, demonstrate the effectiveness and generalization capability of our approach across various object categories and geometric complexities. Moreover, multi-part generation consistently produces meshes with finer geometric details and improved surface quality, highlighting the scalability advantage of our factorized representation for high-resolution mesh synthesis.

\subsection{Ablation Study}
\textbf{On V-VAE.}
Tab.~\ref{tab:v_vae_ablation} studies the main design choices in the vertex branch. Training without procedural synthetic meshes also reduces sparse occupancy accuracy, suggesting that the synthetic set improves coverage of diverse vertex structures. Removing the sub-voxel offset head leaves the sparse occupancy prediction nearly unchanged but noticeably degrades surface reconstruction quality, showing that the offset head specifically improves continuous vertex precision rather than sparse voxel classification. Removing the high-resolution point-to-voxel aggregation and downsampling path causes a larger drop in occupancy metrics and geometric quality, indicating that fine local evidence should be captured before compression to the coarse latent support.

\begin{table}[h]
\centering
\caption{
\textbf{Ablation Study on V-VAE w.r.t. Vertex Reconstruction Performance.}
All metrics are computed between the predicted vertices and ground-truth vertices.
Removing the offset head only affects continuous sub-voxel localization, degrading CD and HD but leaving voxel-level occupancy metrics unchanged.
}
\label{tab:v_vae_ablation}
\begin{tabular}{lcccccc}
\toprule
\textbf{Method} & \textbf{CD(L1)}$\downarrow$ & \textbf{HD}$\downarrow$ & \textbf{ACC}$\uparrow$ & \textbf{F1}$\uparrow$ & \textbf{Recall}$\uparrow$ & \textbf{IOU}$\uparrow$ \\
\midrule
w/o Train on Synthetic Mesh & 0.0005 & 0.0071 & 0.9134 & 0.9137 & 0.9140 & 0.9034 \\
w/o OffsetHead & 0.0011 & 0.0082 & \textbf{0.9752} & \textbf{0.9735} & \textbf{0.9730} & \textbf{0.9603} \\
w/o VDF DownSample & 0.0034 & 0.0145 & 0.8981 & 0.9025 & 0.9069 & 0.8873  \\
\midrule
\textbf{Ours (Full V-VAE)} & \textbf{0.0003} & \textbf{0.0069} & \textbf{0.9752} & \textbf{0.9735} & \textbf{0.9730} & \textbf{0.9603} \\
\bottomrule
\end{tabular}
\end{table}

\textbf{On Topology Generation.}
Tab.~\ref{tab:t_flow_ablation} evaluates the topology generation design. Removing the sparse-structure condition from the topology flow degrades both geometric and topology-related metrics, showing that the shared coarse support provides useful context for generating consistent connectivity. We also compare against a deterministic topology predictor that directly completes edges from vertices without sampling topology latents. The topology latent flow performs better, supporting our claim that mesh connectivity should be modeled as a vertex-conditioned latent distribution rather than a feed-forward completion step.
\begin{table}[h]
\centering
\caption{
\textbf{Ablation study on T-Flow.}
Conditioning T-Flow on the $64$-resolution sparse structure improves final mesh quality, indicating that coarse geometric support helps generate more consistent topology.
}
\label{tab:t_flow_ablation}
\begin{tabular}{lcccc}
\toprule
\textbf{Method} & \textbf{CD(L2)}$\downarrow$ & \textbf{CD(L1)}$\downarrow$ & \textbf{HD}$\downarrow$ & $\boldsymbol{|\mathrm{NC}|}\uparrow$ \\
\midrule
w/o Geometric Condition $\mathbf{c}_\text{g}$
                    & 0.0411 & 0.0601 & 0.0694 & 0.8238 \\
\midrule
\textbf{Ours (Full T-Flow)} 
                    & \textbf{0.0407} & \textbf{0.0596} & \textbf{0.0657} & \textbf{0.8341} \\
\bottomrule
\end{tabular}
\end{table}

\section{Conclusion}

We presented \methodname{}, a factorized sparse latent framework for explicit mesh generation. Instead of representing vertex geometry and mesh topology within a unified latent space, \methodname{} decomposes mesh synthesis into two complementary stages: vertex generation and vertex-conditioned topology generation, each equipped with a dedicated structured latent representation and flow model. The vertex branch reconstructs high-resolution vertice through sparse coarse-to-fine refinement with sub-voxel offset prediction, effectively reducing discretization errors. The topology branch models mesh connectivity as a conditional latent distribution over the generated vertex set, enabling topology synthesis that adapts to arbitrary geometric configurations. Beyond improving mesh generation quality, this factorization enables flexible vertex-level control, including explicit vertex-count manipulation, topology-adaptive mesh editing, and scalable part-wise generation beyond the capacity of monolithic latent representations.

\paragraph{Limitations and future work.}
The current two-stage factorization also introduces several limitations. Since topology generation operates on the generated vertex set, geometric inaccuracies from the vertex stage cannot be corrected during topology synthesis. In addition, topology decoding considers vertex-pair relationships, resulting in quadratic complexity that remains manageable at current resolutions but may require efficient sparse candidate selection for substantially larger meshes. Finally, while \methodname{} focuses on geometry and connectivity generation, production-ready assets typically require additional attributes, such as UV coordinates, textures, and material properties.

Future work will explore tighter interactions between the vertex and topology stages through joint or iterative refinement, enabling topology feedback to improve vertex placement. We also aim to extend the factorized formulation toward comprehensive generation of complete 3D assets with richer appearance attributes. We believe that modeling topology as a conditional latent distribution provides a promising foundation for scalable, controllable, and production-oriented explicit mesh generation.



\bibliography{iclr2026_conference}
\bibliographystyle{iclr2026_conference}



\end{document}